\documentclass{ws-ijmpb}
\usepackage{cite}

\usepackage{epsfig}

\newcommand{\be}{\begin{equation}}
\newcommand{\ee}{\end{equation}}
\newcommand{\ben}{\begin{eqnarray}}
\newcommand{\een}{\end{eqnarray}}

\begin{document}

\markboth{J.~Richter, O.~Derzhko, A.~Honecker}
{The sawtooth chain: From Heisenberg spins to Hubbard electrons}

%
\catchline{}{}{}{}{}
%

\title{The sawtooth chain: From Heisenberg spins to Hubbard electrons}
\author{J.~RICHTER}
\address{Institut f\"ur Theoretische Physik, Otto-von-Guericke
         Universit\"at Magdeburg,\\
         P.O.Box 4120, 39016 Magdeburg, Germany\\
         www.uni-magdeburg.de/itp}

\author{O.~DERZHKO}
\address{Institute for Condensed Matter Physics
         of the National Academy of Sciences of Ukraine,\\
         1 Svientsitskii Street, L'viv-11, 79011, Ukraine}

\author{A.~HONECKER}
\address{Institut f\"ur Theoretische Physik,
         Georg-August-Universit\"at G\"ottingen,\\
         37077 G\"ottingen, Germany}

\maketitle

\begin{history}
\received{\today}
\end{history}

\begin{abstract}
We report on recent studies of the spin-half Heisenberg and the Hubbard
model
on the sawtooth chain. For both models we construct a class of
exact eigenstates
which are localized
due to the frustrating geometry of the lattice
for a certain relation of the exchange (hopping) integrals.
Although  these eigenstates differ in details for the two models
because of the different statistics,
they share some characteristic features.
The localized eigenstates
are highly degenerate and become ground states 
in high magnetic fields (Heisenberg model) 
or at certain electron fillings (Hubbard model), respectively. 
They may dominate
the low-temperature thermodynamics and lead to an extra low-temperature
maximum in the specific heat.
The ground-state degeneracy can be calculated exactly
by a mapping of the manifold of localized ground states
onto a classical hard-dimer problem, and explicit
expressions for thermodynamic quantities can be derived which are
valid at low temperatures near the saturation field for the Heisenberg model 
or around a certain value of the chemical potential for the Hubbard model,
respectively.
\end{abstract}


\section{Introduction}
         \label{secr1}

Frustrated lattices play an important role in the
search for exotic quantum states of condensed matter.
The term `frustration' was introduced in physics in
the 1970s by Toulouse\cite{toulouse}  in the context of spin
glasses\cite{binder} and describes 
a situation where exchange
interactions are in competition with each other. 
The studies on
spin glasses have demonstrated  that frustration
may have an enormous influence
on ground-state and thermodynamic properties
of spin systems\cite{binder}.

In the 1970s Anderson and Fazekas\cite{anderson}
first considered the quantum spin-$1/2$ Heisenberg antiferromagnet
on the geometrically
frustrated triangular
lattice and proposed a liquid-like ground state without magnetic
long-range order.
Although later on it was found
that the spin-$1/2$ Heisenberg antiferromagnet on the triangular
lattice possesses semi-classical three-sublattice N\'{e}el order
(see, e.g., Refs.~\cite{lhuillier01sep,wir04} for recent reviews),
Anderson's suggestion was the starting point to
search for exotic quantum ground states in frustrated spin systems.

The recent progress in synthesizing
frustrated  magnetic materials with strong quantum fluctuations\cite{lemmens}
and the rich behavior of such magnetic systems have stimulated
an enormous interest
in frustrated
quantum magnets,
see, e.g., Refs.~\cite{schiffer,moessner01,phys_today,diep04,002}.
There are many compounds which correspond to quantum antiferromagnetic
Heisenberg models
with frustrated spin interactions. We mention as examples
the frustrated spin-1/2 $J_1-J_2$ chains
(Rb$_{2}$Cu$_{2}$Mo$_{3}$O$_{12}$,
LiCuVO$_4$, Li$_2$ZrCuO$_4$)
\cite{1dj1j2} and
the kagom\'{e} lattice (ZnCu$_3$(OH)$_6$Cl$_2$)\cite{mendels}.
There are also compounds
which correspond to electronic (Hubbard, $t-J$, periodic Anderson) models
on geometrically frustrated lattices.
We mention as examples
cobaltates\cite{cobaltates},
CeRh$_3$B$_2$\cite{pam},
as well as
artificial crystals from quantum dots\cite{tamura+arita}.

\begin{figure}[t]
\centerline{\psfig{file=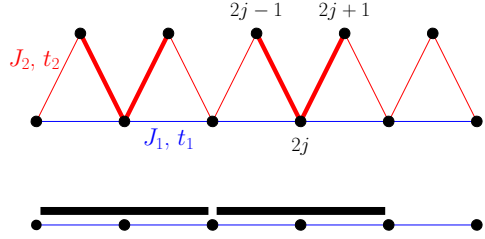,width=2.5in,angle=0}}
\vspace*{8pt}
\caption
{(Color online)
Upper part:
the sawtooth chain. Filled circles indicate the lattice sites,
lines indicate the exchange/hopping bonds.
Two trapping cells occupied by localized magnons/electrons are
indicated by bold lines. Note that for the sawtooth chain
one has two kinds of bonds of different strength.
The lower part of the figure
indicates the corresponding hard-dimer model
(two hard dimers on a linear chain).
}
\label{fig01}
\end{figure}

In this paper we will focus  on a special property
of the Heisenberg and the Hubbard model
on a particular geometrically frustrated lattice (the sawtooth chain, see
Fig.~\ref{fig01}),
namely the existence of
localized eigenstates (on a perfect lattice) and their relevance for the
low-temperature physics of those correlated systems.
Note, however, that arguments and calculations presented in this paper can
in principle be applied to wide class of frustrated lattices, see the
discussion below and
Refs.~\cite{003,004,006,008,009,010,017,018,026,023,024,hub_wir07}.

In general, for perfect lattices an elementary excitation
as a non-interacting quasiparticle is spread
over the entire lattice.
For example, for a simple hypercubic lattice
a  magnon or electron wave function 
is extended over all lattice sites
due to a hopping term in the Hamiltonian.
However,
for some lattice geometries
a wave function of an elementary excitation in a quantum system
may have amplitudes which are non-zero only in a restricted area
owing to destructive quantum interference. We call
such excitations localized excitations
(for example, localized magnons\cite{003,004,006}
or localized electron states\cite{026,023,024,hub_wir07,GKV07,balents}).
Due to the local character of these excitations
exact many-particle eigenstates of the Hamiltonian
can be built by $n$ independent localized excitations
(i.e.\ they have a sufficiently large separation between each other)
even in the presence of interactions.
The number $n$ of localized excitations cannot exceed
a certain maximal value $n_{\max}$ which depends
on the specific
lattice under consideration, where $n_{\max}$
is proportional to the system size $N$\cite{004,hub_wir07}.
If the localized excitation
is the lowest-energy eigenstate of the Hamiltonian
in the one-particle subspace
one may expect
that a state with $n$ independent (isolated)
localized excitations
is the lowest-energy eigenstate of the Hamiltonian
in the corresponding $n$-particle subspace\cite{003,004,005,hub_wir07}
provided there is no attractive interaction.
The localized eigenstates
may become ground states in high magnetic fields (Heisenberg
model) or at certain electron fillings (Hubbard model), respectively.
Therefore they
may substantially contribute to
or even completely dominate
the low-temperature thermodynamic properties of the system.

In the present paper we discuss
the effect of localized elementary excitations on the
low-temperature thermodynamics
focusing on the quantum Heisenberg antiferromagnet and the Hubbard model on
the sawtooth chain.
We follow the lines which have been developed in a series of papers on
localized eigenstates for the
Heisenberg model\cite{003,004,005,006,007,wir04,008,009,010,011,012,013,014,016,017,018,019,020,021,022,022a}
and for electronic models\cite{026,023,024,hub_wir07,GKV07,balents}.

To be specific we consider the 
Heisenberg antiferromagnet of $N$
spins with quantum number $s=1/2$ in a magnetic field $h$
\begin{eqnarray}
H=\sum_{\langle i,j\rangle}
J_{ij}\vec{s}_i\cdot\vec{s}_j
-hS^z
\label{1.01}
\end{eqnarray}
and the Hubbard model of $N$ lattice sites
\begin{eqnarray}
H=\sum_{\langle i,j\rangle \atop \sigma=\uparrow,\downarrow}
t_{ij}\left(c^{\dagger}_{i,\sigma}c_{j,\sigma}+c^{\dagger}_{j,\sigma}c_{i,\sigma}\right)
+U\sum_{i}n_{i,\uparrow}n_{i,\downarrow}
+
\mu\sum_{i,\sigma=\uparrow,\downarrow}n_{i,\sigma}.
\label{hubbard}
\end{eqnarray}
In (\ref{1.01}) and (\ref{hubbard})
the first sum runs over all neighboring sites on the lattice under
consideration,
$J_{ij}>0$ is the antiferromagnetic isotropic Heisenberg exchange interaction
between the sites $i$ and $j$,
and
$S^z=\sum_is_i^z$ is the $z$-component of the total spin.
In (\ref{hubbard})
$t_{ij}>0$ is the hopping matrix element between the nearest-neighbor sites $i$ and $j$,
$U>0$ is the on-site Coulomb repulsion,
$\mu$ is the chemical potential, 
and $n_{i,\sigma}=c^{\dagger}_{i,\sigma}c_{i,\sigma}$.
For electronic models the chemical potential $\mu$ plays the role of the
magnetic
field $h$. While for the Heisenberg antiferromagnet $h$ controls the
magnetization $M=S^z$,
$\mu$ controls the average number of electrons in the system for the Hubbard model.

In what follows we first consider the frustrated quantum
Heisenberg antiferromagnet on the sawtooth chain (Fig.~\ref{fig01})
and
discuss some generic properties
of the model
which are caused by the localized magnon states.
In particular, we consider the magnetization process,
calculate the ground-state
degeneracy of the localized eigenstates leading to a finite residual
entropy and
discuss the low-temperature thermodynamics
for magnetic fields in the vicinity of the saturation field (Sec.~\ref{secr2}).
Then we  illustrate the application of the  concepts elaborated for the spin
system to the Hubbard model on the sawtooth chain in Sec.~\ref{secr3}.
Sec.~\ref{secr4} presents a short summary of our discussion.

\section{Localized Magnon States in the Heisenberg Antiferromagnet on the Sawtooth Chain
         \label{secr2}}
\subsection{Flat bands and localized eigenstates
         \label{secr2_1}}

In this section we illustrate how the localized magnon states emerge
for the frustrated quantum Heisenberg antiferromagnet (\ref{1.01}).
The fact that $S^z$ commutes with the Hamiltonian (\ref{1.01})
permits us to consider the eigenstates separately in each subspace
with different values of $S^z=N/2, N/2 - 1, \ldots $.
In the subspace with $S^z=N/2$
the only eigenstate is the fully polarized ferromagnetic state,
$\vert {\rm{FM}}\rangle=\vert \uparrow\uparrow\uparrow\uparrow\uparrow\ldots\rangle$,
which plays the role of the vacuum state for the magnon excitations.

\begin{figure}[t]
\centerline{\psfig{file=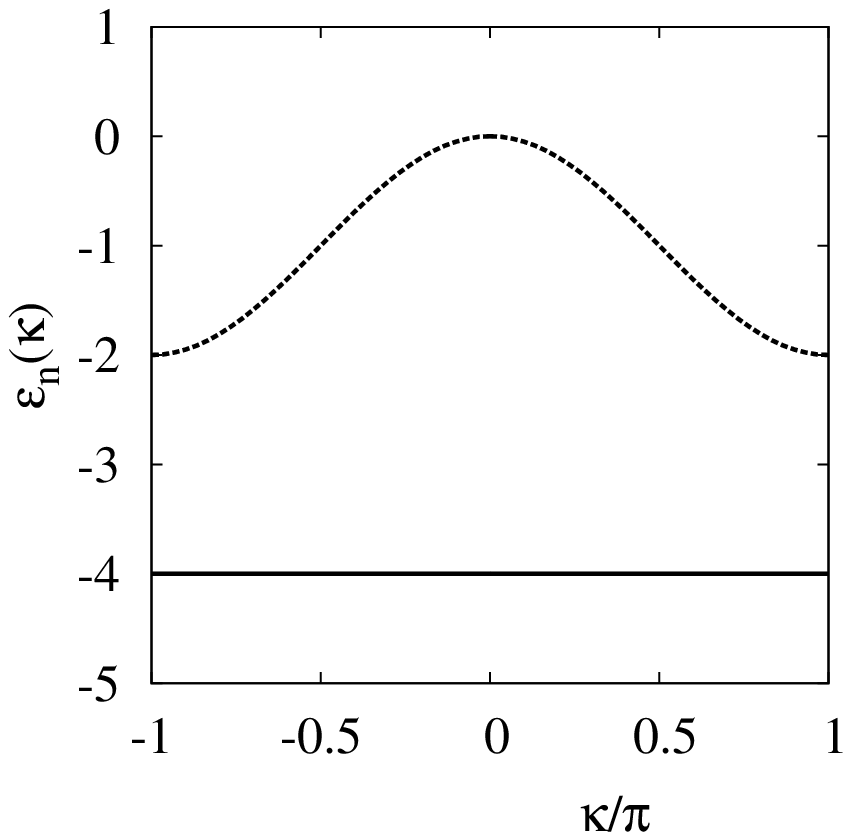,width=2.61in,angle=0}
\psfig{file=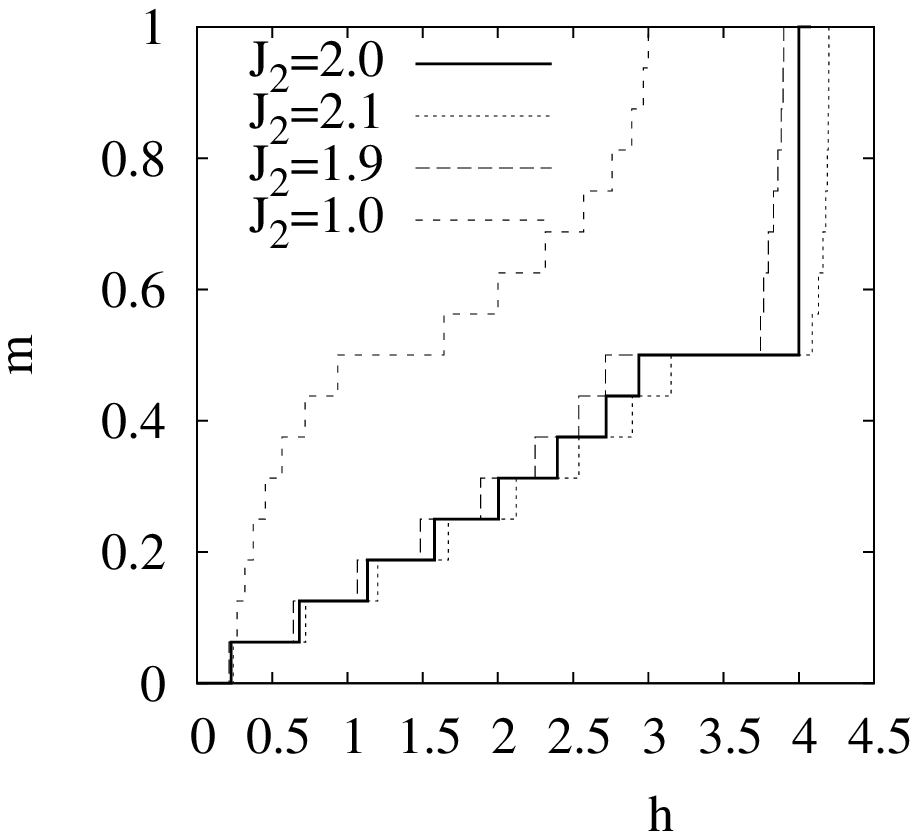,width=2.7in,angle=0}}
\vspace*{8pt}
\caption{Left: One-magnon dispersion for the spin-$1/2$
Heisenberg
antiferromagnet on the sawtooth chain with
$J_2=2$, $J_1=1$  and $h=0$ (cf. Eq.  (\ref{2.05})). 
Right: Ground-state magnetization curves $m(h)=M(h)/M_{\max}$
for the spin-$1/2$ Heisenberg
antiferromagnet on the sawtooth chain for various values of $J_2$ and
$J_1=1$.}
\label{fig02}
\end{figure}

In the one-magnon subspace ($S^z=N/2-1$) it is simple to calculate the eigenstates
given  by
$|1_\kappa\rangle =\sum_{l=0}^1 c_l \sum_{j=1}^{N/2}
{\rm{e}}^{{\rm{i}}\kappa j}s_{2j+l}^-|{\rm{FM}}\rangle$;
$H |1_\kappa\rangle = \varepsilon_{\pm}(\kappa) |1_\kappa\rangle $.
The two one-magnon branches are given by
\be
\varepsilon_{\pm}(\kappa) = h
-\frac{ J_1 + 2J_2 }{2}
+\frac{1}{2} \left[
J_1\cos\kappa  \; \pm
\sqrt{J_1^2\left(-1+\cos\kappa\right)^2+2J_2^2\left(1+\cos
\kappa\right)}\right].
\label{2.05}
\ee
For $J_2=2J_1$,
the lower magnon band becomes completely flat, i.e.\
$\varepsilon_-(\kappa)=\varepsilon_-
=h - 4J_1$, see the left panel of
Fig.~\ref{fig02}. Let us now focus on the case $J_2=2J_1$.
A dispersionless band allows one to construct localized excitations
given here by
$
\vert 1{\rm{lm}}\rangle = l^{\dagger}_{2j} \vert {\rm{FM}} \rangle ,
$
where
$
l^{\dagger}_{2j}=(1/\sqrt{6})\left(s^-_{2j-1}-2s^-_{2j}+s^-_{2j+1}\right)
$
creates a spin excitation (magnon) localized in a valley (trapping cell)
indicated by bold lines in the upper part of Fig.~\ref{fig01}.
Note that a typical geometrical feature of a lattice leading to the
possibility to localize eigenstates is
a triangular configuration of antiferromagnetic bonds, where the triangle is
built  by one bond of the trapping  cell (here a valley)
and two bonds attached to the
trapping cell\cite{004,006}, see Fig.~\ref{fig01}.

Let us consider the $n$-magnon subspace with $S^z=N/2-n$.
In this subspace the construction of the eigenstates of the Heisenberg
model
is, generally,  a difficult many-body problem.
However, for a lattice which supports localized magnon states,
a state $\vert n{\rm{lm}}\rangle$  consisting of $n$ independent
(i.e.\ isolated)
localized magnons
is an exact eigenstate of the Hamiltonian (\ref{1.01}).
Using the  $l^{\dagger}$-operators introduced above
these states can be written as
$\vert n{\rm{lm}}\rangle =
l^{\dagger}_{i_1}l^{\dagger}_{i_2} \ldots l^{\dagger}_{i_n}\vert{\rm{FM}}\rangle$,
where the $i_l$ are sufficiently separated lattice sites.
For the sawtooth chain all $n$ localized magnons are independent
(isolated) if they do not occupy neighboring valleys (hard core rule).
This constraint immediately leads to
a maximum number of localized magnons $n_{\max}=N/4$.
The energy of the $n$-particle state $\vert n {\rm{lm}}\rangle$
is
\be \label{2.15}
E_{n {\rm{lm}}}= E_{{\rm{FM}}}-\frac{N}{2}h + n(h - 4J_1),
\ee
i.e.\ at $h=h_1=4J_1$ all
localized magnon states are degenerate.
It is important to note that the localized magnon states are the lowest
eigenstates in all  sectors of $S^z=N/2-1,N/2-2,\ldots,N/2-n_{\max}$\cite{003,005}.
Hence these states become
ground states in an appropriate magnetic field. Furthermore it can be shown
that all localized magnon states are linearly independent
for the sawtooth chain \cite{019} and
that the localized magnon states
present the complete manifold of ground states in all relevant sectors of
$S^z$\cite{009,010,018}.

In the following sections we will discuss 
how the localized eigenstates influence the physical
properties of  frustrated lattices.

\subsection{Plateaus and jumps in the magnetization curve} \label{m_h}
First we consider the relevance of the localized magnon states for the
magnetization process.
For the calculation of the  magnetization $M=S^z$ at $T=0$
it is sufficient to find
the lowest energy levels $E(M)$ in the subspaces with different 
$M=N/2,N/2-1,\ldots$
for $h=0$.
The energy in the presence of an external magnetic field $h$
is given by
$E(M,h)=E(M)-hM$,
where the magnetization $M$ should acquire a value which minimizes $E(M,h)$.
Hence $M$
can be  determined from the equation
${\rm{d}}E(M) / {\rm{d}}M = h$ which finally
gives the magnetization curve $m(h)$ where $m=M/M_{\max}$, $M_{\max}=N/2$.
For a classical non-frustrated Heisenberg antiferromagnet one typically
finds a parabolic relation
$E(M)\propto M^2$ resulting in a straight-line behavior $M\propto h$.
Often quantum fluctuations
lead only to small deviations from a linear $M-h$ relation,
see, e.g., Refs.~\cite{lhuillier01sep,wir04,honecker04}.
However, in the presence of frustration and quantum fluctuations
more exotic magnetization curves, e.g., curves with plateaus,
can be observed\cite{wir04,lhuillier01sep,honecker04}.
Another spectacular feature observed in magnetization curves of frustrated
quantum spin systems consists in discontinuous jumps
related to a linear relation $E\propto M$\cite{HMT00,003,004,006,wir04,018}.
As discussed in the previous section we find such a linear $E-M$ relation
for the sawtooth Heisenberg antiferromagnet with $J_2=2J_1$ for values of the
magnetization for which the lowest eigenstates are  localized states, see
Eq.~(\ref{2.15}). 
This leads to a magnetization jump from $m=1/2$
directly to saturation $m=1$ at the saturation field $h_1=4J_1$, see
the right panel of Fig.~\ref{fig02}.
In addition there is wide plateau preceding the jump. This plateau state
represents a regular pattern of alternately occupied and empty  valleys and
is two-fold degenerate.
Magnetization curves with a jump to saturation for other lattices can be
found, e.g., in
Refs.~\cite{Mila98,HMT00,003,004,006,honecker04,011,016,021}.
We emphasize that the jump is macroscopic, and that there is no
finite-size effect. Furthermore we mention that a jump to saturation can be found
also for the sawtooth Heisenberg antiferromagnet with higher spin quantum number
$s>1/2$.
However,  the height of the  jump decreases
with $1/s$, i.e.\ the jump is a true quantum effect and disappears in
the classical limit $s \to \infty$.

Finally, let us discuss deviations from the ideal parameter constellation
$J_2=2J_1$ for which the localized magnon states are true
eigenstates. 
The right panel of Fig.~\ref{fig02} 
shows that small deviations (e.g.,
$J_2=1.9J_1$, $J_2=2.1J_1$) do not change the magnetization curve drastically, whereas the
model with uniform bonds $J_2=J_1$ exhibits a qualitatively different $m(h)$
behavior.
\subsection{Ground-state residual entropy and low-temperature
thermodynamics} \label{2.3}

It has been shown above
that the energy of the $n$-magnon state in a magnetic field
is $E_{{\rm{FM}}}-Nh/2 + n(h - 4J_1)$, cf.\ Eq.~(\ref{2.15}).
Obviously, for $n < n_{\max}$
this energy level is highly degenerate,
since there are many ways to place $n$ independent localized magnons on a lattice.
The degeneracy further increases at the saturation field $h_1=4J_1$,
since the energies of the states with different numbers of localized magnons
$n=0,1,\ldots,n_{\max}$ become equal, 
namely $E_{{\rm{FM}}}-Nh_1/2$.
We denote this degeneracy at $h=h_1$ by ${\cal{W}}$.
Since all localized magnon states are linearly independent\cite{019}, they
span a highly degenerate ground-state manifold at $h=h_1$. The
degree of degeneracy can be calculated by taking into account the hard-core
rule (simultaneous occupation of neighboring valleys by localized magnons is
forbidden).
The remaining counting problem can be solved by mapping the localized magnon
problem on the sawtooth chain with $N$ sites onto a hard-dimer problem
(simultaneous occupation of neighboring sites by dimers is forbidden) on a simple linear
chain with ${\cal N}=N/2$ sites, see
the lower part of Fig.~\ref{fig01} and also
Refs.~\cite{008,009,010,018}.
Taking the number of hard-dimer distributions from the
literature\cite{fisher}
we can use this mapping to find the
ground-state degeneracy at the saturation field ${\cal{W}}$.
For ${\cal{N}} \to \infty$  one finds
${\cal{W}}
=
\left((1+\sqrt{5})/2\right)^{{\cal{N}}}
\approx
\exp\left(0.4812 {\cal{N}}\right)$
leading to a finite residual entropy of
$S/k_{\rm{B}}N=(1/2)\ln\left((1+\sqrt{5})/2\right)\approx 0.2406$
for the sawtooth chain with $J_2=2J_1$ at $h=h_1$\cite{008,009,010}.

In addition, we can use the correspondence
between the localized magnon states and the spatial configurations
of hard dimers
to calculate the contribution of the localized magnon states
to the thermodynamic quantities following the lines given, e.g., in
Refs.~\cite{fisher,035}.
This contribution may dominate the
low-temperature thermodynamics
and therefore we may find predictions
for the low-temperature behavior of the magnetic quantities
in the vicinity of the saturation
field $h_1$.
The contribution of the localized states to the
partition function of the spin model
can be written as
\begin{eqnarray}
Z_{{\rm{lm}}}(T,h,N)
&=&\exp\left(-\frac{E_{{\rm{FM}}}-h\frac{N}{2}}{k_{\rm{B}}T}\right)
\sum_{n=0}^{n_{\max}} g_N(n)\exp\left(\frac{h_1-h}{k_{\rm{B}}T}n\right)
\nonumber\\
&=&\exp\left(-\frac{E_{{\rm{FM}}}-h\frac{N}{2}}{k_{\rm{B}}T}\right)\Xi(T,\mu,{\cal{N}})
\; ; \quad {\cal N} = \frac{N}{2}.
\label{2.19}
\end{eqnarray}
Here $g_N(n)$ is the degeneracy of the ground state of the spin model
with $N$ sites in the sector with $n$ localized magnons, i.e.\ with
$M=S^z=N/2-n$.
In the hard-dimer description $g_N(n)$
corresponds to the canonical partition function $Z(n,{\cal{N}})$
of the classical hard-dimer model.
$h_1-h=\mu$ is the chemical potential of the hard dimers
and $\Xi(T,\mu,{\cal{N}})$ (or $\Xi(z,{\cal{N}})$,
$z=\exp\left(\mu/k_{\rm{B}}T\right)$)
is the grand-canonical partition function of the one-dimensional hard-dimer lattice gas
given by
\be \label{grand}
\Xi(T,\mu,{\cal{N}})
=\lambda_1^{{\cal{N}}}+\lambda_2^{{\cal{N}}},
\;\;\;
\lambda_{1,2}=\frac{1}{2}\pm\sqrt{\frac{1}{4}+\exp x},
\;\;\;
x=\frac{\mu}{k_{\rm{B}}T} .
\ee
Formula (\ref{2.19}) describes the low-temperature thermodynamics
of the spin model near the saturation field accurately,
i.e.\ $Z(T,h,N) \approx Z_{{\rm{lm}}}(T,h,N)$,
because of the huge degeneracy of the ground state at $h=h_1$ (note that
there are no other ground states apart from
the considered localized-magnon states in the corresponding sectors of
$S^z$).
We mention that similar considerations are possible for other frustrated
lattices~\cite{009,010,017,018,020,022,022a}.
The contribution of the localized magnon states
to the Helmholtz free energy $F$ of the spin model
is given by
$F_{{\rm{lm}}}(T,h,N)/N
=E_{{\rm{FM}}}/N - h/2  -k_{\rm{B}}T
\ln\Xi(z,{\cal{N}})/N$.
The entropy $S$,
the specific heat $C$,
the magnetization $M$
and the susceptibility $\chi$ follow from $F_{{\rm{lm}}}(T,h,N)$
according to usual relations
$S_{{\rm{lm}}}(T,h,N)=-\partial F_{{\rm{lm}}}(T,h,N)/\partial T$,
$C_{{\rm{lm}}}(T,h,N)=T\partial S_{{\rm{lm}}}(T,h,N)/\partial T$,
$M_{{\rm{lm}}}(T,h,N)=N/2-\langle n\rangle
=N/2-k_{\rm{B}}T\partial \ln\Xi(T,\mu,{\cal{N}})/\partial\mu$,
$\chi_{{\rm{lm}}}(T,h,N)=\partial M_{{\rm{lm}}}(T,h,N)/\partial h$.
In the limit $N\to\infty$ this leads to \cite{008,017,018}
\begin{eqnarray}
\nonumber\\
\frac{S_{{\rm{lm}}}(T,h,N)}{k_{\rm{B}}N}
&=&\frac{1}{2}\left[\ln\left(\frac{1}{2}+\sqrt{\frac{1}{4}+\exp x}\right)
-x\left(\frac{1}{2}-\frac{1}{4\sqrt{\frac{1}{4}+\exp x}}\right)\right],
\nonumber\\
\frac{C_{{\rm{lm}}}(T,h,N)}{k_{\rm{B}}N}
&=&\frac{1}{16}
\frac{x^2\exp x}{\left(\frac{1}{4}+\exp x\right)^{\frac{3}{2}}}, \nonumber
\\
\frac{M_{{\rm{lm}}}(T,h,N)}{\frac{N}{2}}
&=&1-
\left(\frac{1}{2}-\frac{1}{4\sqrt{\frac{1}{4}+\exp x}}\right),
\nonumber\\
\frac{k_{\rm{B}}T\chi_{{\rm{lm}}}(T,h,N)}{N}
&=&\frac{1}{16}
\frac{\exp x}{\left(\frac{1}{4}+\exp x\right)^{\frac{3}{2}}} \; ; \quad
x=\frac{\mu}{k_{\rm{B}}T}=\frac{h_1-h}{k_{\rm{B}}T}.
\label{2.23}
\end{eqnarray}
The thermodynamic quantities depend on $T$ and $h$ via the
universal parameter  $x =(h_1-h)/k_{\rm{B}}T$ only.
Corresponding formulas for finite systems can be found using
$\Xi(T,\mu,{\cal{N}})$ from Eq.~(\ref{grand})
in combination with the relation between $F_{{\rm{lm}}}(T,h,N)$
and $\Xi(T,\mu,{\cal{N}})$ given above.

\begin{figure}[t]
\centerline{\psfig{file=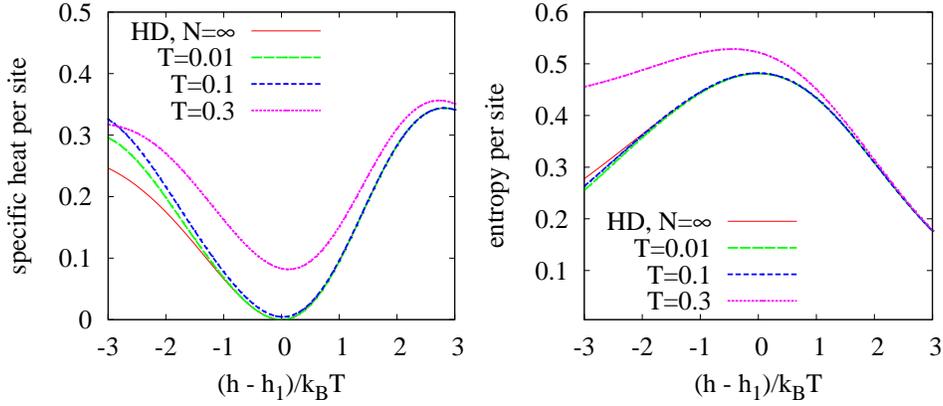,width=5.0in,angle=0}
}
\vspace*{8pt}
\caption
{(Color online) The specific heat (left)
and the entropy (right) in dependence  on $x=(h_1-h)/k_{\rm{B}}T$
for the spin-$1/2$ Heisenberg antiferromagnet
on the sawtooth chain with $N=20$, $J_1=1$, $J_2=2$ and $k_BT=0.01, 0.1,
0.3$ in comparison with
the one-dimensional hard-dimer (HD) gas, Eq.~(\ref{2.23}).
}
\label{fig06}
\end{figure}

\begin{figure}[t]
\centerline{\psfig{file=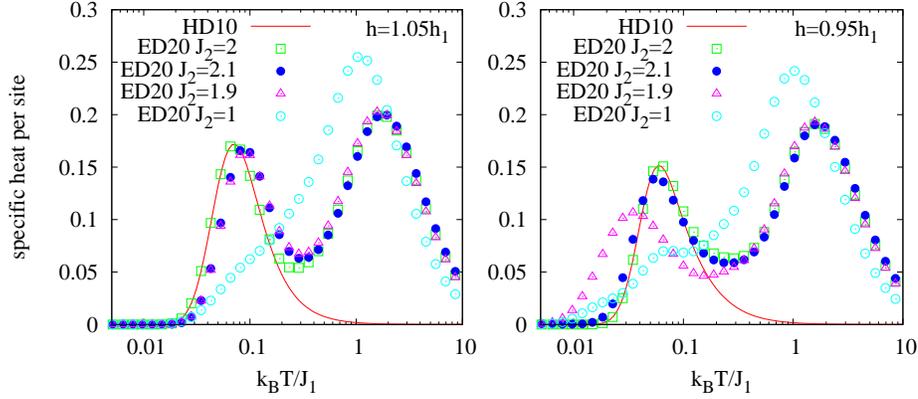,width=5.3in,angle=0}}
\vspace*{8pt}
\caption
{(Color online) Temperature dependence of the specific heat for the
spin-$1/2$ Heisenberg antiferromagnet
on the sawtooth chain with $N=20$, $J_1=1$ and various $J_2$ for magnetic
fields slightly above ($h=1.05h_1$, left) and below ($h=0.95h_1$, right)
the saturation field
($h_1 = 4$, $4.2$, $3.9$ and $3$
for $J_2=2$, $2.1$, $1.9$ and $1$, respectively). For comparison we show
the results for the one-dimensional hard-dimer (HD) gas with ${\cal N}=10$
sites.
}
\label{fig06a}
\end{figure}

Fig.~\ref{fig06} shows a comparison of
the entropy and the specific heat of the spin model in dependence on the
universal parameter $x=(h_1-h)/k_{\rm{B}}T$ with the
hard-dimer formulas.
In addition we show the specific heat as an important measurable
quantity in dependence on the temperature for magnetic fields slightly
above and below  the saturation field in Fig.~\ref{fig06a}.

We emphasize here some prominent features:
an extra low-temperature peak in the dependence $C$ vs.\ $T$ for fields
slightly below or slightly above $h_1$ (Figs.~\ref{fig06} and
\ref{fig06a})
and
an enhanced entropy at $h_1$ at low temperatures (Fig.~\ref{fig06}). Note
that $C$ in Eq.~(\ref{2.23}) is zero at $x=0$ and consequently there is no extra peak
in $C(T)$ for $h=h_1$, see also Fig.~\ref{fig06} (left).
Furthermore from Figs.~\ref{fig06} and \ref{fig06a} it becomes evident
that the hard-dimer description
works excellently for temperatures up to 10\% of the exchange
coupling and reproduces qualitatively
the characteristic features of the spin model for
higher temperatures up to about $0.3J_1$.

Similar as for the  magnetization curve we consider now
the influence of deviations from the ideal parameter constellation
$J_2=2J_1$ (for which the localized magnon states are true exact
eigenstates) on thermodynamic quantities.
From Fig.~\ref{fig06a} it is obvious that only large deviations suppress
the extra low-temperature peak in $C(T)$.
This behavior can be explained by inspection of
the low-energy spectrum. For small deviations the energy is only
slightly changed and the originally highly degenerate ground-state manifold
becomes quasi-degenerate. As a result the $\delta$-peak present in the
low-energy density of
states for $J_2=2J_1$ is broadened but there is still a well pronounced
maximum in the density of states leading to the extra low-$T$ peak in
$C(T)$.

Let us very briefly discuss an aspect of the localized magnon scenario which
might have some relevance for a possible application of highly frustrated
magnets.
Due to the huge degeneracy of the localized magnon states and the resulting
residual entropy at $h=h_1$ there is a well pronounced  low-temperature peak
in the entropy $S$ versus field $h$ curve, see Fig.~\ref{fig06} (right).
It has been pointed out first by Zhitomirsky\cite{Zhi:PRB03}
considering the classical
kagom\'{e} Heisenberg antiferromagnet that such a degeneracy leads to an
enhanced magnetocaloric
effect.  Later on  this point has been discussed for quantum spin systems,
e.g., in Refs.~\cite{008,017,018,schnack}.

\section{Hubbard Electrons on the Sawtooth Chain
         \label{secr3}}
\subsection{Flat one-electron band and localized electron eigenstates}

We consider now the Hubbard model (\ref{hubbard}) on a sawtooth chain. The
specific Hamiltonian reads
\begin{eqnarray}
H
&=&\sum_{j=0}^{\frac{N}{2}-1}\sum_{\sigma=\uparrow,\downarrow}
\Bigg [
t_1c_{2j,\sigma}^{\dagger}c_{2j+2,\sigma}
+t_2\Big(c_{2j,\sigma}^{\dagger}c_{2j+1,\sigma}
+c_{2j+1,\sigma}^{\dagger}c_{2j+2,\sigma}\Big)
+{\rm{h.c.}}
\nonumber\\
&& +\mu \left(n_{2j,\sigma}
+n_{2j+1,\sigma}
\right)\Bigg ]
+U\sum_{j=0}^{\frac{N}{2}-1}
\left(n_{2j,\uparrow}n_{2j,\downarrow}
+n_{2j+1,\uparrow}n_{2j+1,\downarrow}
\right),
\label{3.01}
\end{eqnarray}
where $t_1>0$ and $t_2>0$ are the hopping integrals
along the base line and the zig-zag path, respectively (see
the upper part of Fig.~\ref{fig01}),
and $U>0$ is the on-site Coulomb repulsion.
The sawtooth-chain Hubbard model has attracted much attention since the
1990s\cite{penc}.
Here we focus on a special aspect, namely the existence of localized ground
states and their consequences for the low-temperature physics of the model.
On the one-particle level
the description of the electron system
is the same as of the $XY$ spin system\cite{023,024,hub_wir07}.
The one-electron dispersion reads
\be \varepsilon_{\pm}(\kappa)
=\mu
+t_1\cos\kappa
\pm
\sqrt{t_1^2\cos^2\kappa+2t_2^2\left(1+\cos \kappa\right)}\;\;.
\ee
Thus,
if $t_2=\sqrt{2} t_1$
the lowest single electron energy becomes $\varepsilon_-=\mu-2t_1$, i.e.\ it is completely flat.
Similar as for the Heisenberg model we can construct
$N$ localized one-electron  ground states, given by
$l^{\dagger}_{2j,\sigma}\vert 0\rangle$,
$l^{\dagger}_{2j,\sigma}
=(1/2)(c^{\dagger}_{2j-1,\sigma}-\sqrt{2}c^{\dagger}_{2j,\sigma}+c^{\dagger}_{2j+1,\sigma})$
(i.e.\ the electron is localized in any of
the $N/2$ valleys labeled by the index $2j$ and
having either spin up or spin down)
with energy $\varepsilon_-=-2t_1+\mu$. Note that the indices of the
$l^{\dagger}$ and $c^{\dagger}$ operators correspond to the lattice sites as
illustrated in Fig.~\ref{fig01}.

The Hubbard repulsion becomes relevant in the two-electron subspace.
Obviously, a two-particle ground state can be constructed
by two independent localized electrons with arbitrary spin
trapped on two valleys which do not touch each other.
However, in contrast to the Heisenberg model there is no `hard-core rule',
i.e.\ there are further two-particle ground states with  two electrons trapped on two
neighboring valleys, e.g., with indices $2j$ and $2j+2$.
The energy of the corresponding eigenstates
$ l_{2j,\uparrow}^{\dagger} l_{2j+2,\uparrow}^{\dagger} \vert 0\rangle$
and
$ l_{2j,\downarrow}^{\dagger} l_{2j+2,\downarrow}^{\dagger} \vert 0\rangle$
 is also independent of $U$, since both electrons have
the same spin  and therefore the Pauli principle
forbids the simultaneous occupation of the site $2j+1$ belonging to both
valleys.
In addition, a straightforward direct calculation shows that for two electrons having different spin
the  linear combination
\begin{eqnarray}
l_{2j,\uparrow}^{\dagger}l_{2j+2,\downarrow}^{\dagger}\vert 0\rangle
+
l_{2j,\downarrow}^{\dagger}l_{2j+2,\uparrow}^{\dagger}\vert 0\rangle,
\label{3.02}
\end{eqnarray}
is also a ground state in the two-electron subspace with an energy
independent of $U$.
This can be seen also  by using the SU(2) symmetry of the Hubbard
Hamiltonian:
the state (\ref{3.02}) and the states
$l_{2j,\uparrow}^{\dagger} l_{2j+2,\uparrow}^{\dagger} \vert 0\rangle$
and
$l_{2j,\downarrow}^{\dagger} l_{2j+2,\downarrow}^{\dagger} \vert 0\rangle$
form a triplet, i.e.\ (\ref{3.02}) can be obtained by acting
with the total spin lowering operator 
$S^-=\sum_i c_{i,\downarrow}^{\dagger}c_{i,\uparrow}$
on the state
$l_{2j,\uparrow}^{\dagger} l_{2j+2,\uparrow}^{\dagger} \vert
0\rangle$.
Of course,
all states belonging to one triplet have the same energy $2(\mu-2t_1)$.

We can generalize this procedure to construct the ground states in the subspaces
with $n=3,\ldots,N/2$ electrons
\be \label{n_states}
|\varphi^{\uparrow}_{n}
\rangle \propto {  l^\dagger_{2i_n,\uparrow } \cdots
l^\dagger_{2i_1,\uparrow}}
\vert 0 \rangle \quad ; \quad
{H}|\varphi^{\uparrow}_{n}\rangle =
n(-2t_1+\mu)|\varphi^{\uparrow}_{n}\rangle.
\ee
They are all
degenerate for $\mu=\mu_0=2t_1$ and do not feel $U$.
Evidently, they are fully polarized
\be {S}^z|\varphi^{\uparrow}_{n}\rangle =
 \frac{n}{2}|\varphi^{\uparrow}_{n}\rangle
\quad ; \quad
{\vec  S}^2|\varphi^{\uparrow}_{n}\rangle =
 \frac{n}{2}\left(\frac{n}{2}+1\right)|\varphi^{\uparrow}_{n}\rangle .
\ee
Again
the application of $S^-$ yields new eigenstates with the same energy
and the
same ${\vec S}^2$,
but
with
$S^z (S^-)^k|\varphi^{\uparrow}_{n} \rangle = (n/2 - k)(S^-)^k|\varphi^{\uparrow}_{n} \rangle$.
Note that  ${ l^\dagger_{2i_n,\uparrow }} \cdots
{ l^\dagger_{2i_k,\downarrow }} \cdots
{ l^\dagger_{2i_1,\uparrow}}
\vert 0 \rangle$, where $i_1,\ldots,i_k,\ldots,i_n$ denote $n$ contiguous
valleys, is not an
eigenstate.
Since there is no hard-core rule
the maximum filling with localized electrons  is $n_{\max} = N/2$, i.e.\ it is twice as
large as for localized magnons.

In the next step we use the 
fully polarized $n$-electron states
$|\varphi^{\uparrow}_{n}\rangle$ to construct the complete
set of ground states for $0\le n \le N/2$. The
$|\varphi^{\uparrow}_{n}\rangle$
can be grouped into two classes, namely in
one-cluster states
and in  multi-cluster states. While for the one-cluster states the
electrons occupy a cluster of contiguous valleys,
for a multi-cluster state
the electrons occupy two or more clusters, where each cluster  is built by
contiguous valleys and the different clusters are separated by one or more
empty valleys. The key observation is that
 further ground states can be constructed by application of a certain
cluster spin flip operator $S^-_{\rm clust}=\sum_{i \in {\rm clust}}
c_{i,\downarrow}^{\dagger}c_{i,\uparrow}$ on a multi-cluster $n$-electron ground state $|\varphi^{\uparrow}_{n}
\rangle$. The resulting new states are not fully polarized and
complete the set of ground states in each sector $n$\cite{hub_wir07}.

\subsection{Hole concentration in dependence on the  chemical
potential}\label{n_mu}
\begin{figure}[t]
\centerline{\psfig{file=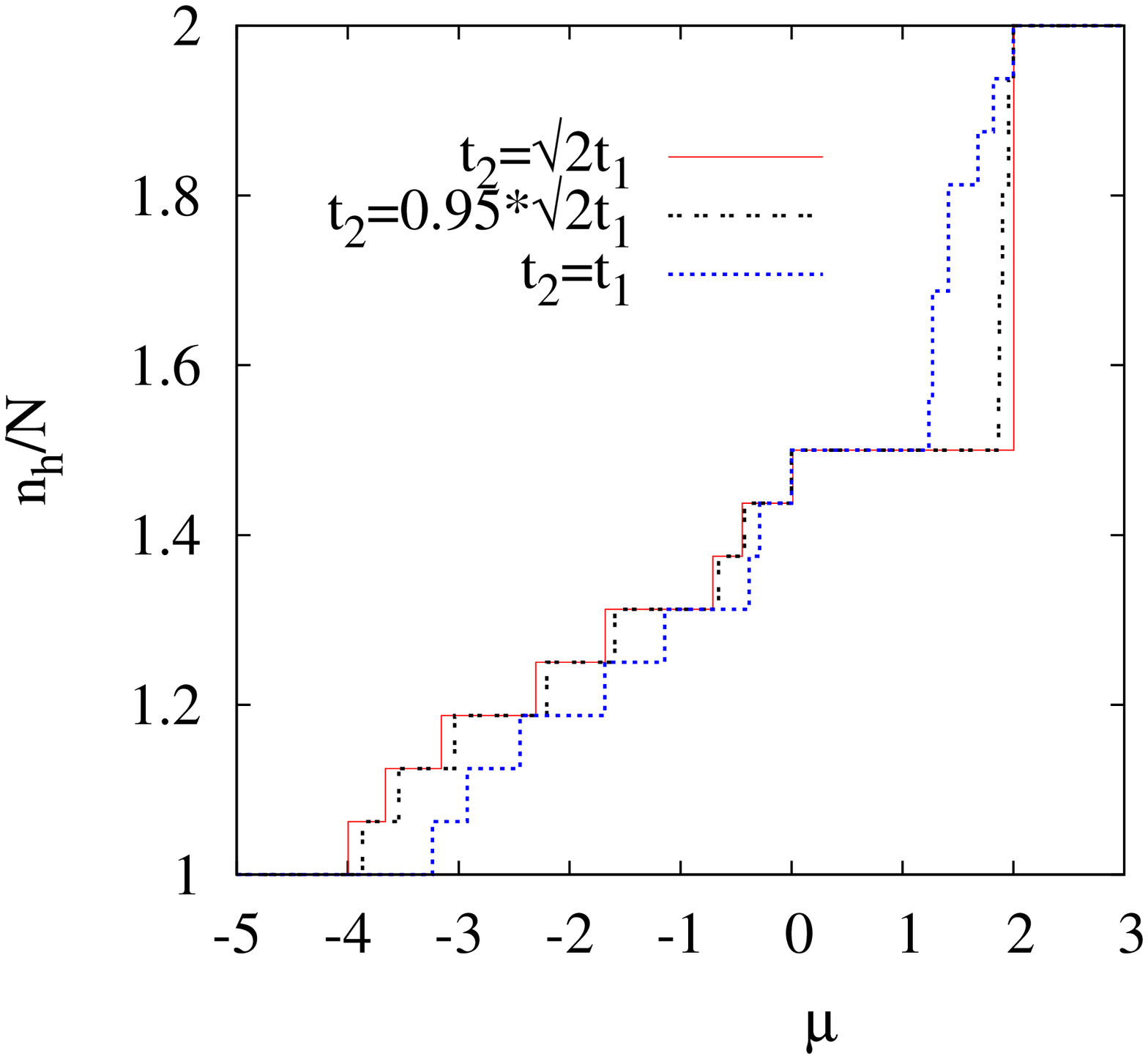,width=2.2in,angle=270}
\psfig{file=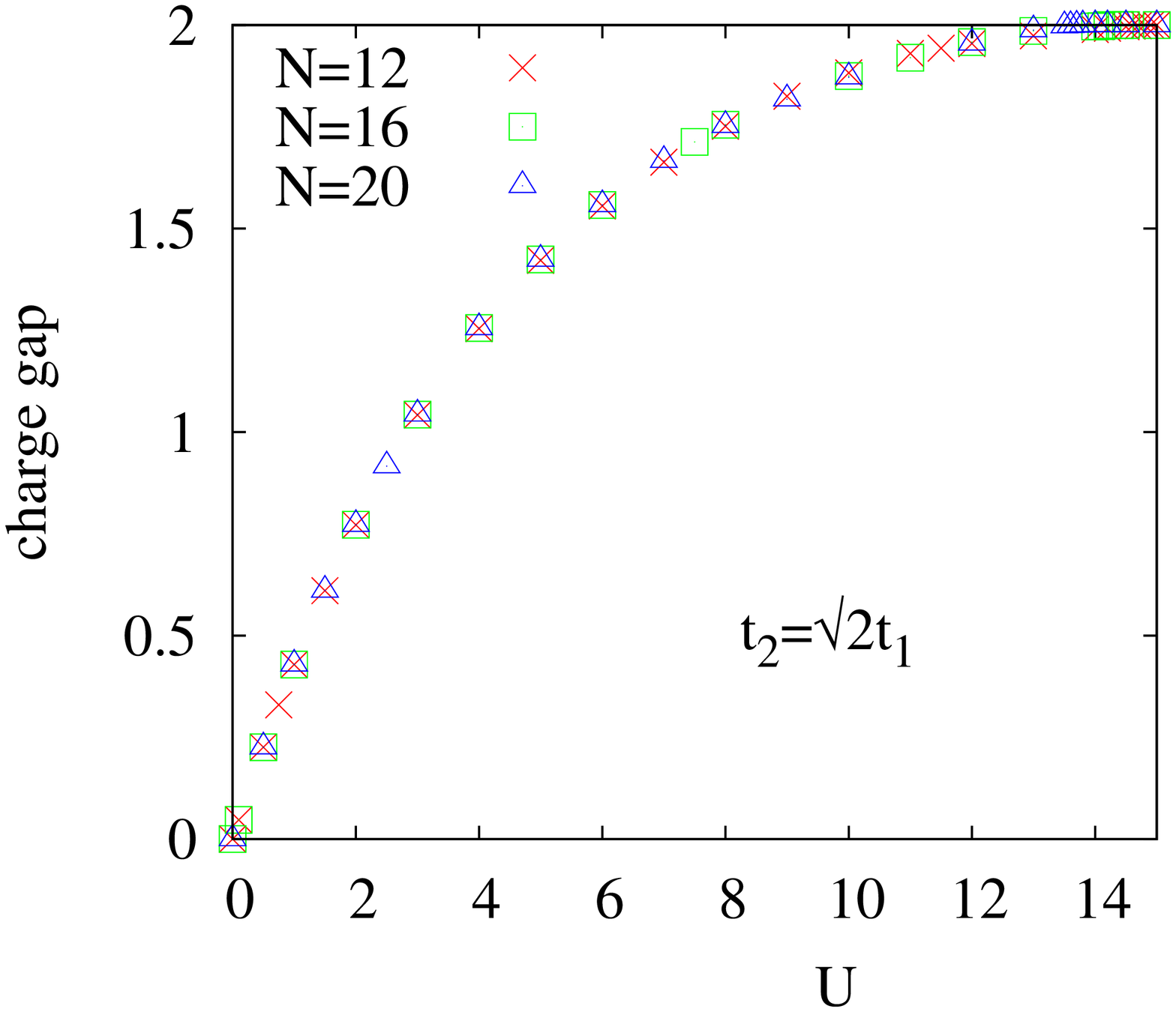,width=2.2in,angle=270}}
\vspace*{8pt}
\caption{(Color online) Left: Hole concentration $n_{\rm h}/N=2-n/N$ versus
chemical potential $\mu$ for $t_2=\sqrt{2}t_1$ (localized-electron
regime), $t_2=0.95\sqrt{2}t_1$ and $t_2=t_1$ for a finite sawtooth
chain of $N=16$
sites (periodic boundary conditions) and $U\to \infty$, $t_1=1$.
Right: Charge gap $\Delta \mu = E(N/2+1)-2\,E(N/2)+E(N/2-1)$
at quarter filling versus $U$ for $t_2=\sqrt{2}t_1$, $t_1=1$ and $N=12,16,20$.}
\label{fig08}
\end{figure}

In correspondence to the $m(h)$ curve of the Heisenberg model we consider
now the hole concentration $n_{\rm h}/N=2-n/N$ in dependence on the chemical
potential $\mu$ (Fig.~\ref{fig08} (left)).
Like for spin systems, see Sec.~\ref{m_h},
the main characteristics for the system with localized eigenstates (i.e.\ for
$t_2=\sqrt{2}t_1$) are a size-independent jump of $n_{\rm h}/N$
from $3/2$ to $2$ and a plateau at $n_{\rm h}/N=3/2$.
This plateau determines the range of validity
of the localized-electron picture at $T=0$.
The right panel of Fig.~\ref{fig08} presents the
plateau width, i.e.\ the size of the charge gap,
versus $U$ for $N=12$, $16$, and $20$.
One observes
that there is almost no finite-size dependence. Since the charge gap
is zero for $U=0$ and increases with $U$ we conclude that its appearance is due
to the on-site repulsion.
Small deviations from the ideal parameter values $t_2=\sqrt{2}t_1$
do not change the $n_{\rm h}/N$ versus $\mu$ curve substantially,
as illustrated for the case $t_2=0.95\sqrt{2}t_1$ in Fig.~\ref{fig08} (left),
whereas for the
model with uniform hopping integrals $t_2=t_1$ the charge gap is
significantly smaller and there is no indication of a jump from the
plateau at $n_{\rm h}/N=3/2$ to $n_{\rm h}/N=2$.

\subsection{Ground-state residual entropy and low-temperature thermodynamics}

The localized-electron states are {\em linearly independent},
which is connected with the fact
(as in the case of spin systems, see Ref.~\cite{019}) that
the middle site is unique to each valley.
Therefore all these highly degenerate states contribute to the partition
function.
Now the question arises whether the ground state degeneracy
can
be calculated.
Due to the different statistics of Hubbard electrons and Heisenberg spins there
are some differences
in the construction rules of localized eigenstates (e.g., the occupation of
neighboring valleys is forbidden for spins but allowed for electrons, see
above).
Hence it is not surprising that
the ground state degeneracy $g_N(n)$ for $n$ electrons on the $N$-site
sawtooth chain does not coincide with the one
for the Heisenberg sawtooth chain
(which was equal to the canonical partition functions
of $n$ hard dimers on a simple chain of ${\cal{N}}=N/2$ sites, see Sec.~\ref{2.3}).
Nevertheless,
$g_N(n)$, $n=0,1,2,\ldots,N/2$
for the Hubbard sawtooth chain can also be found by a mapping of the
localized-electron degrees of freedom onto
the one-dimensional hard-dimer problem.
However, this mapping is more intricate and
hard dimers have to be considered
on a simple chain of $N$ sites
(instead of $N/2$ sites as for Heisenberg spins), for details see
Ref.~\cite{hub_wir07}.
One finds
$g_N(n)=Z(n,N)$ for $n=0,1,\ldots,N/2-1$
and
$g_{N}(N/2)=N/2+1=Z(N/2,N)+N/2-1$ where $Z(n,N)$ is the canonical
partition function of the classical one-dimensional hard-dimer
model\cite{fisher,035}.
As for spin systems we can calculate
the contribution of localized electron states to the
partition function by using this mapping.
Again  we can present analytical formulas for the
low-temperature thermodynamic quantities for a non-trivial quantum many-body
problem.
The grand-canonical partition function $\Xi$ of the electron system for
a chemical potential $\mu$ in the vicinity of  $\mu_0=2t_1$
takes the form
\begin{eqnarray}
\Xi(T,\mu,N)
&=&\lambda_1^N+\lambda_2^N+\lambda_3^N,
\nonumber\\
\lambda_{1,2}
&=&\frac{1}{2}\pm\sqrt{\frac{1}{4}+\exp x},
\;\;\;
\lambda_3=\left(\frac{N}{2}-1\right)^{\frac{1}{N}}\exp\frac{x}{2},\;\;\;
x=\frac{2t-\mu}{k_{\rm{B}}T}.
\label{3.03}
\end{eqnarray}
In the thermodynamic limit $N\to\infty$
only the largest eigenvalue $\lambda_1$
of the transfer matrix survives
and, using the definitions
$S(T,\mu,N) = k_B\partial\left(T\, \ln \Xi(T,\mu,N)\right)/\partial T$,
$C(T,\mu,N) = T\,\partial S(T,\mu,N)/\partial T$,
we obtain the following results for
the thermodynamics of one-dimensional hard dimers
(see also \cite{hub_wir07})
\begin{eqnarray}
\frac{S(T,\mu,N)}{k_{\rm{B}}N}
&=&\ln\left(\frac{1}{2}+\sqrt{\frac{1}{4}+\exp x}\right)
-x\left(\frac{1}{2}-\frac{1}{4\sqrt{\frac{1}{4}+\exp x}}\right),
\nonumber\\
\frac{C(T,\mu,N)}{k_{\rm{B}}N}
&=&\frac{x^2\exp x}{8\left(\frac{1}{4}+\exp x\right)^{\frac{3}{2}}},
\nonumber\\
\frac{\langle n\rangle}{N}
&=&\frac{1}{2}-\frac{1}{4\sqrt{\frac{1}{4}+\exp x}},
\label{3.04}
\end{eqnarray}
which are quite similar to the corresponding expressions
for Heisenberg spins,
see Eq.~(\ref{2.23}).
Again we have a finite residual entropy
$S/k_{\rm{B}}N=\ln((1+\sqrt{5})/2) \approx 0.4812$, which is twice as large as for the
Heisenberg model.

\begin{figure}[t]
\centerline{\psfig{file=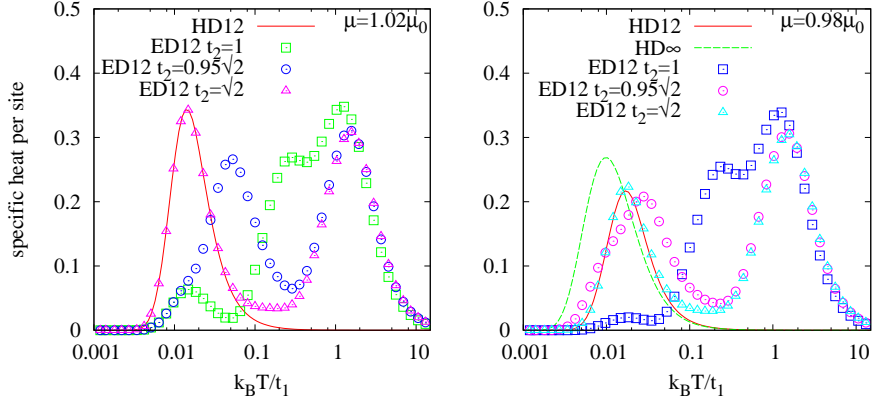,width=5.3in,angle=0}}
\vspace*{8pt}
\caption
{(Color online) Grand-canonical
specific heat per site $C(T,\mu,N)/k_{{\rm{B}}}N$ vs.\ temperature
for the sawtooth Hubbard chain of $N=12$ sites for two values of $\mu$,
$U=\infty$
and $t_1=1$, $t_2=\sqrt{2}$, $0.95\sqrt{2}$ and $1$ (symbols).
Note that $\mu_0 = 2$
for $t_2=\sqrt{2}$, $0.95\sqrt{2}$ and $1$.
For comparison we show the hard-dimer data
for $N=12$ (solid line) which follows from Eq. (\ref{3.03}) and for
$N=\infty$ (dashed line, Eq.~(\ref{3.04})). Note that for $\mu=1.02\mu_0$
the hard-dimer data for $N=12$ and $N=\infty$ practically coincide.
}
\label{fig09}
\end{figure}

Results for the low-temperature
grand-canonical specific heat are shown in
Fig.~\ref{fig09} for two values of the
chemical potential slightly above and below $\mu_0$.
Similar as for the spin system we see (i) that the
hard-dimer model, Eqs.~(\ref{3.03}) and (\ref{3.04}), yields a good description
of the electronic model at low temperatures and (ii) that there is an extra
low-temperature maximum in the grand-canonical specific heat due to
the manifold of localized electron ground states. Again this
additional
low-temperature maximum in $C(T)$ disappears at $\mu=\mu_0$ as 
can be
read off from Eq.~(\ref{3.04}) (note that $C(x=0)=0$).

At the end of this section we would like to mention a relation
to the so-called flat-band ferromagnetism in the Hubbard
model found by Mielke
and Tasaki in the early 1990s\cite{026}. In particular, the ground states
belonging to the plateau at $n=N/2$, see Sec.~\ref{n_mu}, are fully polarized
ferromagnetic states.
For further details of flat-band ferromagnetism in the sawtooth-chain
Hubbard model the interested reader is
referred to the original papers of Tasaki\cite{026} but also to
Ref.~\cite{hub_wir07}.

\section{Summary
         \label{secr4}}

To summarize,
we have illustrated  some basic concepts of localized eigenstates
in correlated systems on highly frustrated lattices
and their effect on the low-temperature thermodynamics.
As a rule non-interacting electrons or magnons on a lattice are delocalized,
i.e.\ are described by a wave function distributed over the whole lattice.
Electrons or magnons  may become localized due to randomness
or after switching on interactions.
As we have discussed on this paper, a
frustrating lattice topology may lead to another
mechanism for localization.
Localized states may survive in the presence of interactions
and under certain conditions they can determine the properties of the system
at low temperatures.

\section*{Acknowledgments}
\addcontentsline{toc}{section}{Acknowledgments}

The authors would like to thank J.~J\c{e}drzejewski,
T.~Krokhmalskii, R.~Moessner, H.-J.~Schmidt,
J.~Schnack, J.~Schulenburg
and M.~E.~Zhitomirsky for useful discussions and
fruitful collaboration in this field.
A.~H.\ acknowledges financial support by the Deutsche Forschungsgemeinschaft
through a Heisenberg fellowship (grant HO~2325/4-1).
We mention that most of the numerical results presented in
this article were obtained using
J.~Schulenburg's {\it spinpack}.


\begin{thebibliography}{000}

\bibitem{toulouse}
G.~Toulouse,
{\it Commun.~Phys.} {\bf 2}, 115 (1977).

\bibitem{binder}
K.~Binder and A.~P.~Young,
{\it Rev.~Mod.~Phys.} {\bf 58}, 801 (1986).

\bibitem{anderson}
P.~W.~Anderson,
{\it Mater.~Res.~Bull.} {\bf 8}, 153 (1973);
P.~W.~Anderson and P.~Fazekas,
{\it Phil.~Mag.}  {\bf 30}, 423 (1974).

\bibitem{lhuillier01sep}
 C.~Lhuillier and G.~Misguich,
in: {\it High magnetic fields},
C.~Berthier, L.P.~L\'evy, G.~Martinez, Eds.
(Lecture Notes in Physics, {\bf 595}) 
(Sprin\-ger, Berlin, 2001),
pp. 161-190.

\bibitem{wir04}
J.~Richter, J.~Schulenburg and A.~Honecker,
in:
{\it Quantum Magnetism},
U.~Schollw\"{o}ck, J.~Richter, D.~J.~J.~Farnell, R.~F.~Bishop, Eds.
(Lecture Notes in Physics, {\bf 645})
(Sprin\-ger, Berlin, 2004), pp. 85-153.

\bibitem{lemmens}
P.~Lemmens and P.~Millet,
in:
{\it Quantum Magnetism},
U.~Schollw\"{o}ck, J.~Richter, D.~J.~J.~Farnell, R.~F.~Bishop, Eds.
(Lecture Notes in Physics, {\bf 645})
(Sprin\-ger, Berlin, 2004), pp. 433-477.

\bibitem{schiffer}
P.~Schiffer,
{\it Nature} {\bf 413}, 48 (2001).

\bibitem{moessner01}
R.~Moessner,
{\it Can.~J.~Phys.} {\bf 79}, 1283 (2001).

\bibitem{phys_today}
R.~Moessner and A.~P.~Ramirez,
{\it Physics Today}, February 2006, p. 24.

\bibitem{diep04}
{\it Frustrated Spin Systems},
H.~T.~Diep, Ed.
(World Scientific, Singapore, 2004).

\bibitem{002}
{\it Quantum Magnetism},
U.~Schollw\"{o}ck, J.~Richter, D.~J.~J.~Farnell, R.~F.~Bishop, Eds.
(Lecture Notes in Physics, {\bf 645})
(Sprin\-ger, Berlin, 2004).

\bibitem{1dj1j2}
M.~Hase, H.~Kuroe, K.~Ozawa, O.~Suzuki, H.~Kitazawa, G.~Kido and T.~Sekine,
{\it Phys.~Rev.~B} {\bf 70}, 104426 (2004);
M.~Enderle, C.~Mukherjee, B.~F{\aa}k, R.~K.~Kremer, J.-M.~Broto,
H.~Rosner, S.-L.~Drechsler, J.~Richter, J.~Malek, A.~Prokofiev,
W.~Assmus, S.~Pujol, J.-L.~Raggazzoni, H.~Rakoto, M.~Rheinst\"{a}dter
and H.~M.~R{\o}nnow,
{\it Europhys.~Lett.} {\bf 70}, 237 (2005);
M.~G.~Banks, F.~Heidrich-Meisner, A.~Honecker, H.~Rakoto,
J.-M.~Broto and R.~K.~Kremer,
{\it J.~Phys.: Condens.~Matter} {\bf 19}, 145227 (2007);
S.-L.~Drechsler, O.~Volkova, A.~N.~Vasiliev, N.~Tristan,
J.~Richter, M.~Schmitt, H.~Rosner, J.~M\'{a}lek, R.~Klingeler,
A.~A.~Zvyagin and B.~B\"{u}chner,
{\it Phys.~Rev.~Lett.} {\bf 98}, 077202 (2007).

\bibitem{mendels}
P.~Mendels, F.~Bert, M.~A.~de~Vries, A.~Olariu, A.~Harrison, F.~Duc,
J.~C.~Trombe, J.~S.~Lord, A.~Amato and C.~Baines,
{\it Phys.~Rev.~Lett.} {\bf 98}, 077204 (2007).

\bibitem{cobaltates}
K.~Takada, H.~Sakurai, E.~Takayama-Muromachi, F.~Izumi, R.~A.~Dilanian and T.~Sasaki,
{\it Nature} {\bf 422}, 53 (2003);
Y.~Wang, N.~S.~Rogado, R.~J.~Cava and N.~P.~Ong,
{\it Nature} {\bf 423}, 425 (2003);
M.~L.~Foo, Y.~Wang, S.~Watauchi, H.~W.~Zandbergen, T.~He, R.~J.~Cava and N.~P.~Ong,
{\it Phys.~Rev.~Lett.} {\bf 92}, 247001 (2004).

\bibitem{pam} 
H.~N.~Kono and Y.~Kuramoto,
{\it J.~Phys.~Soc.~Jpn.} {\bf 75}, 084706 (2006).

\bibitem{tamura+arita}
H.~Tamura, K.~Shiraishi, T.~Kimura and H.~Takayanagi,
{\it Phys.~Rev.~B} {\bf 65}, 085324 (2002);
R.~Arita, K.~Kuroki, H.~Aoki, A.~Yajima, M.~Tsukada,
S.~Watanabe, M.~Ichimura, T.~Onogi and T.~Hashizume,
{\it Phys.~Rev.~B} {\bf 57}, R6854 (1998).

\bibitem{003}
J.~Schnack, H.-J.~Schmidt, J.~Richter and J.~Schulenburg,
{\it Eur.~Phys.~J.~B} {\bf 24}, 475 (2001).

\bibitem{004}
J.~Schulenburg, A.~Honecker, J.~Schnack, J.~Richter and H.-J.~Schmidt,
{\it Phys.~Rev.~Lett.} {\bf 88}, 167207 (2002).

\bibitem{006}
J.~Richter, J.~Schulenburg, A.~Honecker, J.~Schnack and H.-J.~Schmidt,
{\it J.~Phys.: Condens.~Matter} {\bf 16}, S779 (2004).

\bibitem{008}
M.~E.~Zhitomirsky and A.~Honecker,
{\it J.~Stat.~Mech.:~Theor.~Exp.}, P07012 (2004).

\bibitem{009}
M.~E.~Zhitomirsky and H.~Tsunetsugu,
{\it Phys.~Rev.~B} {\bf 70}, 100403(R) (2004).

\bibitem{010}
O.~Derzhko and J.~Richter,
{\it Phys.~Rev.~B} {\bf 70}, 104415 (2004).

\bibitem{017}
M.~E.~Zhitomirsky and H.~Tsunetsugu,
{\it Prog.~Theor.~Phys.~Suppl.} {\bf 160}, 361 (2005).

\bibitem{018}
O.~Derzhko and J.~Richter,
{\it Eur.~Phys.~J.~B} {\bf 52}, 23 (2006).

\bibitem{026}
A.~Mielke,
{\it J.~Phys.~A} {\bf 24}, L73 (1991);
A.~Mielke,
{\it J.~Phys.~A} {\bf 24}, 3311 (1991);
A.~Mielke,
{\it J.~Phys.~A} {\bf 25}, 4335 (1992);
H.~Tasaki,
{\it Phys.~Rev.~Lett.} {\bf 69}, 1608 (1992);
A.~Mielke and H.~Tasaki,
{\it Commun.~Math.~Phys.} {\bf 158}, 341 (1993);
H.~Tasaki,
{\it Prog.~Theor.~Phys.} {\bf 99}, 489 (1998).

\bibitem{023}
A.~Honecker and J.~Richter,
{\it Condensed Matter Physics (L'viv)} {\bf 8}, 813 (2005).

\bibitem{024}
A.~Honecker and J.~Richter,
{\it J.~Magn.~Magn.~Mater.} {\bf 310}, 1331 (2007).

\bibitem{hub_wir07}
O.~Derzhko, A.~Honecker and J.~Richter,
{\it Phys.~Rev.~B} {\bf 76}, 220402(R) (2007).

\bibitem{GKV07}
Z. Gul\'acsi, A. Kampf and D. Vollhardt,
{\it  Phys.~Rev.~Lett.} {\bf 99}, 026404 (2007).

\bibitem{balents}
C. Wu, D. Bergman, L. Balents, and S.~Das~Sarma,
{\it  Phys.~Rev.~Lett.} {\bf 99}, 070401 (2007);
D.~L.~Bergman, C.~Wu, and L.~Balents, {\it  Phys.~Rev.~B} {\bf 78}, 125104
(2008).


\bibitem{005}
H.-J.~Schmidt,
{\it J.~Phys.~A} {\bf 35}, 6545 (2002).

\bibitem{007}
J.~Richter, O.~Derzhko and J.~Schulenburg,
{\it Phys.~Rev.~Lett.} {\bf 93}, 107206 (2004).

\bibitem{011}
J.~Richter, J.~Schulenburg, A.~Honecker and D.~Schmalfu{\ss},
{\it Phys.~Rev.~B} {\bf 70}, 174454 (2004).

\bibitem{012}
J.~Richter, J.~Schulenburg, P.~Tomczak and D.~Schmalfu{\ss},
arXiv:cond-mat/0411673.

\bibitem{013}
R.~Schmidt, J.~Richter and J.~Schnack,
{\it J.~Magn.~Magn.~Mater.} {\bf 295}, 164 (2005).

\bibitem{014}
O.~Derzhko and J.~Richter,
{\it Phys.~Rev.~B} {\bf 72}, 094437 (2005).

\bibitem{016}
J.~Richter,
{\it Fizika Nizkikh Temperatur (Kharkiv)} {\bf 31}, 918 (2005)
[{\it Low Temperature Physics} {\bf 31}, 695 (2005)].

\bibitem{019}
H.-J.~Schmidt, J.~Richter and R.~Moessner,
{\it J.~Phys.~A} {\bf 39}, 10673 (2006).

\bibitem{020}
J.~Richter, O.~Derzhko and T.~Krokhmalskii,
{\it Phys.~Rev.~B} {\bf 74}, 144430 (2006);
O.~Derzhko, J.~Richter and T.~Krokhmalskii,
{\it Acta Physica Polonica A} {\bf 113}, 433 (2008).

\bibitem{021}
J.~Schnack, H.-J.~Schmidt, A.~Honecker, J.~Schulenburg and J.~Richter,
{\it J.~Phys.: Conf.~Ser.} {\bf 51}, 43 (2006).

\bibitem{022}
O.~Derzhko, J.~Richter, A.~Honecker and H.-J.~Schmidt,
{\it Fizika Nizkikh Temperatur (Kharkiv)} {\bf 33}, 982 (2007)
[{\it Low Temperature Physics} {\bf 33}, 745 (2007)].

\bibitem{022a}
M.~E.~Zhitomirsky and H.~Tsunetsugu,
{\it Phys.~Rev.~B} {\bf 75}, 224416 (2007).

\bibitem{honecker04}
A.~Honecker, J.~Schulenburg and J.~Richter,
{\it J.~Phys.: Condens.~Matter} {\bf 16}, S749 (2004).

\bibitem{HMT00}
A.~Honecker, F.~Mila and M.~Troyer,
{\it Eur.~Phys.~J.~B} {\bf 15}, 227 (2000).

\bibitem{Mila98}
F.~Mila,
{\it Eur.~Phys.~J.~B} {\bf 6}, 201 (1998).


\bibitem{fisher}
M.~E.~Fisher,
{\it Phys.~Rev.} {\bf 124,} 1664 (1961).

\bibitem{035}
R.~J.~Baxter,
{\it Exactly Solved Models in Statistical Mechanics}
(Academic Press, London, 1982).

\bibitem{Zhi:PRB03}
M.~E.~Zhitomirsky,
{\it Phys.~Rev.~B} {\bf 67}, 104421 (2003).

\bibitem{schnack}
J.~Schnack, R.~Schmidt and J.~Richter,
{\it Phys.~Rev.~B} {\bf 76}, 054413 (2007).

\bibitem{penc}
K.~Penc, H.~Shiba, F.~Mila and T.~Tsukagoshi,
{\it Phys.~Rev.~B} {\bf 54}, 4056 (1996);
H.~Sakamoto and K.~Kubo,
{\it J.~Phys.~Soc.~Jpn.} {\bf 65}, 3732 (1996);
Y.~Watanabe and S.~Miyashita,
{\it J.~Phys.~Soc.~Jpn.} {\bf 66}, 2123;
Y.~Watanabe and S.~Miyashita,
{\it J.~Phys.~Soc.~Jpn.} {\bf 66}, 3981 (1997).

\end{thebibliography}
\end{document}